\documentclass[12pt]{article}

\title{A Classical Explanation of the Bohm-Aharonov Effect}
\author{O. Chavoya-Aceves\\Camelback HS\\Phoenix AZ USA\\E-mail chavoyao@yahoo.com}
\date{}

\begin{document}
\maketitle

\abstract{The motion of a system of particles under
electromagnetic interaction is considered. Under the assumption
that the force acting on an electric charge is given by the sum of
the electromagnetic fields produced by any other charged particles
in its neighborhood, we prove that the vector potential of the
electromagnetic field has to be considered for the balance of
kinetic momentum. The theory cannot be quantized in the usual
form---because it involves a mass matrix that depends on spatial
variables---and the Hamilton's function becomes singular at a
distance equal to the geometric mean of the electrodynamic
radiuses of electrons and protons.}

\section{Introduction}
In previous works \cite{chavoya-03} \cite{chavoya-04}, we have
shown that, according to classical mechanics and electrodynamics:
a neutral system of electric charges that passes through a region
where there is an inhomogeneous magnetic field, experiences a
force, even if its internal \emph{kinetic} angular momentum is
equal to zero. Given that this challenges the common
interpretation of the Stern-Gerlach experiment---as evidence that
there are intrinsic angular momenta---we have considered necessary
to study the motion of systems of electric charges where the
internal magnetic force is not neglected, as it is usual in common
classical treatments.

In this paper we study the motion of electric charges under
electromagnetic interaction. Neglecting radiative effects, we
assume that the force acting on an electric charge is given by the
sum of the electromagnetic fields produced by any other particles
in its neighborhood. From the invariance of the Lagrange's
function, we find that the vector potential of the electromagnetic
field must be considered for the balance of linear and angular
momentum, thus predicting a classical \emph{Bohm-Aharonov Effect}.

A Hamilton's function is obtained also for the system of two
particles. The result is a theory that cannot be quantized but
approximately in the usual form, since it involves a mass matrix
that depends on spatial variables. Also, the Hamilton's function
becomes singular where the distance between the particles
satisfies the relation:
\[
r=\frac{e^2}{c^2(m_em_p)^{1/2}}.
\]

For the sake of completeness, we include a section where the
Lagrange's function and the equations of motion for the center of
mass and the vector of relative position are obtained.

\section{The General Law of Motion}
We study the classical motion of an electron and a proton, under
electromagnetic interaction.

Neglecting any retardation and/or radiative effects, we use the
formulas
\begin{equation}\label{electrodynamic potentials}
  \phi(\vec{x},t)=\frac{q}{\|\vec{x}-\vec{r}(t)\|}\ and\
  \vec{A}(\vec{x},t)=\frac{q}{c}\frac{\vec{v}(t)}{\|\vec{x}-\vec{r}(t)\|},
\end{equation}
to find the electrodynamic potentials associated to a punctual
charge $q$ moving along the path $\vec{r}(t)$. (Where
$\vec{v}=\frac{d\vec{r}}{dt}$.) The corresponding electromagnetic
field is
\begin{equation}\label{electric field}
  \vec{E}(\vec{x},t)=-\nabla\phi-\frac{1}{c}\frac{\partial \vec{A}}{\partial
  t}=
\end{equation}
\[
\frac{q(\vec{x}-\vec{r})}{\|\vec{x}-\vec{r}\|^3}-
\frac{q}{c^2}\left(\frac{\dot{\vec{v}}}{\|\vec{x}-\vec{r}\|}+%
 \frac{((\vec{x}-\vec{r})\cdot\vec{v})\vec{v}}{\|\vec{x}-\vec{r}\|^3}\right)
\]
\begin{equation}\label{magnetic field}
  \vec{H}(\vec{x},t)=\nabla\times\vec{A}=\frac{q}{c}\frac{\vec{v}\times (\vec{x}-\vec{r})}%
  {\|\vec{x}-\vec{r}\|^3}.
\end{equation}

Further, we suppose---as it's done when only Coulomb's field is
considered---that the force that acts on the electron is that due
to the proton's electromagnetic field, and \emph{vice versa}. The
equations of motion are:
\begin{equation}\label{ecuacion para el electron}
  m_e\dot{\vec{v}_e}=-\frac{e^2(\vec{r}_e-\vec{r}_p)}{r^3}%
  +\frac{e^2}{c^2}\left(\frac{\dot{\vec{v}_p}}{r}+\frac{((\vec{r}_e-\vec{r}_p)\cdot\vec{v}_p)\vec{v}_p}{r^3}
  \right)
\end{equation}
\[
-\frac{e^2}{c^2}\frac{\vec{v}_e\times(\vec{v}_p\times(\vec{r}_e-\vec{r}_p))}{r^3}
\]
and
\begin{equation}\label{ecuacion para el proton}
  m_p\dot{\vec{v}_p}=-\frac{e^2(\vec{r}_p-\vec{r}_e)}{r^3}%
  +\frac{e^2}{c^2}\left(\frac{\dot{\vec{v}_e}}{r}+\frac{((\vec{r}_p-\vec{r}_e)\cdot\vec{v}_e)\vec{v}_e}{r^3}
  \right)
\end{equation}
\[
-\frac{e^2}{c^2}\frac{\vec{v}_p\times(\vec{v}_e\times(\vec{r}_p-\vec{r}_e))}{r^3}
\]
(Here we have introduced the notation
\begin{equation}\label{definicion de r}
  r=\|\vec{r}_p-\vec{r}_e\|
\end{equation}
that simplifies the equations.)

These are the Euler-Lagrange's Equations for the Lagrange's
Function:
\begin{equation}\label{lagrange function}
  L(\vec{r}_e,\vec{r}_p,\vec{v}_e,\vec{v}_p)=%
  \frac{1}{2}m_e\vec{v}_e^{\ 2}+\frac{1}{2}m_p\vec{v}_p^{\ 2}+%
  \frac{e^2}{r}-\frac{e^2}{c^2}%
  \frac{\vec{v}_e\cdot\vec{v}_p}{r},
\end{equation}
as we'll prove for the equation
\begin{equation}\label{euler lagrange proton}
  \frac{d}{dt}\frac{\partial L}{\partial \vec{v}_p}-%
  \frac{\partial L}{\partial\vec{r}_p}=\vec{0}.
\end{equation}

From \ref{lagrange function} we get
\begin{equation}\label{derivada del impulso del pronton}
    \frac{d}{dt}\frac{\partial L}{\partial \vec{v}_p}=%
    m_p\dot{\vec{v}_p}-\frac{e^2}{c^2}\frac{d}{dt}\left(%
\frac{\vec{v}_e}{r}
    \right).
\end{equation}
Further:
\begin{equation}\label{ecuacion auxiliar uno}
  \frac{d}{dt}\left(\frac{\vec{v}_e}{r}\right)=%
  \frac{\dot{\vec{v}_e}}{r}-\frac{((\vec{r}_p-\vec{r}_e)\cdot\vec{v}_p)\vec{v}_e}{r^3}
  +\frac{((\vec{r}_p-\vec{r}_e)\cdot\vec{v}_e)\vec{v}_e}{r^3},
\end{equation}
and
\begin{equation}\label{ecuacion auxiliar dos}
  \frac{\partial L}{\partial \vec{r}_p}=-\frac{e^2(\vec{r}_p-\vec{r}_e)}%
  {r^3}+\frac{e^2}{c^2}\frac{(\vec{v}_e\cdot\vec{v}_p)(\vec{r}_p-\vec{r}_e)}{r^3}
\end{equation}
From equations \ref{euler lagrange proton}, \ref{ecuacion auxiliar
uno}, and \ref{ecuacion auxiliar dos}---and the identity
$\vec{a}\times(\vec{b}\times\vec{c})=(\vec{a}\cdot\vec{c})\vec{b}-(\vec{a}\cdot\vec{b})\vec{c}$---
 we can easily prove that equation \ref{euler lagrange proton} is
equivalent to equation \ref{ecuacion para el proton}.

The generalized momenta are
\begin{equation}\label{impulso del electron}
  \vec{p}_e=m_e\vec{v}_e-\frac{e^2}{c^2}\frac{\vec{v}_p}{r}=%
  m_e\vec{v}_e-\frac{e}{c}\vec{A}_p(\vec{r}_e)
\end{equation}
and
\begin{equation}\label{impulso del proton}
  \vec{p}_p=m_p\vec{v}_p-\frac{e^2}{c^2}\frac{\vec{v}_e}{r}=%
  m_p\vec{v}_p+\frac{e}{c}\vec{A}_e(\vec{r}_p)
\end{equation}
where $\vec{A}_e(\vec{r})$ and $\vec{A}_p(\vec{r})$ are the vector
potentials for the field of the electron and the field of the
proton, respectively.

The Lagrange's function \ref{lagrange function} is invariant under
translations and rotations of the reference system; therefore, the
sum of the generalized momenta
\begin{equation}\label{impulso total}
  \vec{p}_e+\vec{p}_p=m_e\vec{v}_e+m_p\vec{v}_p-\frac{e}{c}\vec{A}_p(\vec{r}_e)%
   +\frac{e}{c}\vec{A}_e(\vec{r}_p),
\end{equation}
and the total angular momentum (which is not equal to the
\emph{kinetic momentum} and, therefore, the dipolar field is not
enough to describe the magnetic properties of the system)
\begin{equation}\label{angular momentum}
\vec{L}=
  \vec{r}_e\times\vec{p}_e+\vec{r}_p\times\vec{p}_p,
\end{equation}
are constants of motion. In consequence, the center of mass of the
system,
\begin{equation}\label{center of mass}
  \vec{R}=\frac{m_e\vec{r}_e+m_p\vec{r}_p}{m_e+m_p},
\end{equation}
does not move according to Newton's First Law. This was expected
given that equations \ref{ecuacion para el electron} \&
\ref{ecuacion para el proton} are not in compliance with Newton's
Third Law either.

In the case of a system of $n$ particles with masses
$m_1,\cdots,m_n$ and charges $q_1,\cdots,q_n$, the Lagrange's
Function assumes the form
\begin{equation}\label{funcion de lagran muchas particulas}
  L=\sum_{i=1}^n \frac{1}{2}m_i\vec{v}_i^{\ 2}-\frac{1}{2}\sum_{q_i\ne
  q_j}\frac{q_iq_j}{r_{ij}}\left(1-\frac{\vec{v}_i\cdot\vec{v}_j}{c^2}\right).
\end{equation}
Therefore, the generalized momentum of the $ith$ particle is
\begin{equation}\label{momento de la iesima particula}
  \vec{p}_i=m_i\vec{v}_i+\frac{q_i}{c}\sum_{j\ne
  i}\frac{q_j}{c}\frac{\vec{v}_j}{r_{ij}}=%
  m_i\vec{v}_i+\frac{q_i}{c}\vec{A}_i(\vec{r_i}),
\end{equation}
where $\vec{A}_i$ is the vector potential of the magnetic field
produced by the other particles.

Given that the function \ref{funcion de lagran muchas particulas}
is also invariant under arbitrary translations and/or rotations,
we come again to the conclusion that the sum of the generalized
momenta and the angular momentum, are constants of motion.
Therefore, as it has been confirmed by Bohm and
Aharonov\cite{bohm-aharonov}, the vector potential of the
electromagnetic field acting on each particle must be considered
for the balance of linear momentum.

The problem of gauge invariance is not an issue for us. Under the
gauge transformation
\begin{equation}\label{gauge transformation}
  L'= L+\frac{\partial \lambda}{\partial
  t}+\sum_{i=1}^n\vec{v}_i\cdot\frac{\partial \lambda}{\partial \vec{r}_i}
\end{equation}
the momenta are transformed as:
\begin{equation}\label{momenta gauge transformation}
  {\vec{p}_i}\ '= \vec{p}_i+\frac{\partial \lambda}{\partial
  \vec{r}_i};
\end{equation}
If the sum of the momenta is going to be a constant of motion,
$\lambda$ must be invariant under arbitrary translations. In other
words
\begin{equation}\label{condicion sobre lambda}
  \sum_{i=1}^n\frac{\partial \lambda}{\partial \vec{r}_i}=\vec{0},
\end{equation}
and, in those circumstances:
\[
\sum_{i=1}^n\vec{p}_i\ '=\sum_{i=1}^n\vec{p}_i.
\]

Going back to the electron-proton system, its energy
\begin{equation}\label{energy}
  E=\frac{1}{2}m_e\vec{v}_e^{\ 2}+\frac{1}{2}m_p\vec{v}_p^{\
  2}-\frac{e^2}{c^2}\frac{\vec{v}_p\cdot\vec{v}_e}{r}%
  -\frac{e^2}{r}
\end{equation}
\[
=\frac{1}{2}(\vec{p_e}\cdot\vec{v_e}+\vec{p}_p\cdot\vec{v}_p)-\frac{e^2}{r},
\]
is also a constant of motion.

Solving \ref{impulso del electron} and \ref{impulso del proton}
for the velocities, we find
\begin{equation}\label{velocidad del electron}
  \vec{v}_e=\frac{\frac{\vec{p}_e}{m_e}-\frac{e^2\vec{p}_p}{m_em_pc^2r}}{1-\frac{e^4}{m_em_pc^4r^2}}
\end{equation}
\begin{equation}\label{velocidad del proton}
  \vec{v}_p=\frac{\frac{\vec{p}_p}{m_p}-\frac{e^2\vec{p}_e}{m_em_pc^2r}}{1-\frac{e^4}{m_em_pc^4r^2}}
\end{equation}

From this and \ref{energy} we get the Hamilton's Function
\begin{equation}\label{funcion de hamilton}
  H=\frac{\frac{\vec{p}_e^{\ 2}}{2m_e}+\frac{\vec{p}_p^{\
  2}}{2m_p}-\frac{e^2\vec{p}_e\vec{p}_p}{m_em_pc^2r}}{1-\frac{e^4}{m_em_pc^4r^2}}-\frac{e^2}{r},
\end{equation}

Where
\begin{equation}\label{region of validity}
  r>>\frac{e^2}{c^2 (m_em_p)^{1/2}}
\end{equation}
this function coincides with the function used in \cite{HONGLIANG}
to approximate the eigenvalues of the corresponding quantum
system.

Further, we notice that  \ref{funcion de hamilton} is singular
where
\begin{equation}\label{distance of failure}
  r=\frac{e^2}{c^2 (m_em_p)^{1/2}}.
\end{equation}
At this distance the equations of motion \ref{ecuacion para el
electron} and \ref{ecuacion para el proton} cannot be solved for
the accelerations, which is fundamental for the applicability of
the theorem of existence and uniqueness of solutions. Therefore,
even if the Principle of Least Action is valid, to determine a
particular solution additional conditions have to be imposed.
There are another two possibilities which we shall not investigate
further in this paper:
\begin{enumerate}
\item That inequality \ref{region of validity} defines the
limits of validity of electrodynamics.
\item That a fully relativistic approach is required. In
this case the effects of retardation have to be considered; the
equations of motion are difference-differential equations; and
there is not room for a variational approach. (At least not for a
variational approach that does not explicitly accounts for the
action of the entire electromagnetic field.)
\end{enumerate}

\section{Separation of the Internal Motion}
As it was shown before, the center of mass of the system does not
move according to Newton's First Law. Notwithstanding, and for the
sake of completeness, we'll carry out the decomposition of the
motion into the motion of the center of mass and an internal
motion. Let's consider the substitutions:
\begin{equation}\label{introduccion del centro de masas}
  \vec{R}=\frac{m_p\vec{r}_p+m_e\vec{r}_e}{M};\ \ \
  \vec{r}=\vec{r}_e-\vec{r}_p
\end{equation}
(where $M=m_p+m_e$), in such way that:
\begin{equation}\label{sustitucion de la coordenada del proton}
  \vec{r}_p=\vec{R}-\frac{m_e}{M}\vec{r};%
\ \ \ \vec{r}_e=\vec{R}+\frac{m_p}{M}\vec{r}.
\end{equation}
and
\[
\vec{v}_p\cdot\vec{v}_e=\dot{\vec{R}}^2+K_L\dot{\vec{R}}\cdot\dot{\vec{r}}%
-\frac{m_em_p}{M^2}\dot{\vec{r}}^2,
\]
where
\[
K_L=\frac{m_p-m_e}{M}.
\]

 The Lagrange's function \ref{lagrange function} takes the form:
\begin{equation}\label{lagrangiana en el centro de masas}
  L(\vec{R},\vec{r},\dot{\vec{R}},\dot{\vec{r}})=\frac{1}{2}M\dot{\vec{R}}^2%
  +\frac{1}{2}\mu\dot{\vec{r}}^2+\frac{e^2}{r}-\frac{e^2}{c^2}\frac{\dot{\vec{R}}^2+K_L\dot{\vec{R}}\cdot\dot{\vec{r}}%
-\frac{m_em_p}{M^2}\dot{\vec{r}}^2}{r}.
\end{equation}

The momenta are:
\begin{equation}\label{momento del centro de masas}
  \vec{P}_{\vec{R}}=M\left(1-\frac{2e^2}{Mc^2r}\right)\dot{\vec{R}}-\frac{K_Le^2}{c^2r}\dot{\vec{r}},
\end{equation}
and
\begin{equation}\label{momento del movimiento interno}
  \vec{p}_{\vec{r}}=-\frac{K_Le^2}{c^2r}\dot{\vec{R}}+\mu\left(1+\frac{2e^2}{M
  c^2r}\right)\dot{\vec{r}}.
\end{equation}

The energy is:
\begin{equation}\label{energia}
  E(\vec{R},\vec{r},\dot{\vec{R}},\dot{\vec{r}})=\frac{1}{2}M\dot{\vec{R}}^2%
  +\frac{1}{2}\mu\dot{\vec{r}}^2-\frac{e^2}{r}-\frac{e^2}{c^2}\frac{\dot{\vec{R}}^2+K_L\dot{\vec{R}}\cdot\dot{\vec{r}}%
-\frac{m_em_p}{M^2}\dot{\vec{r}}^2}{r}
\end{equation}
\[
=\frac{1}{2}(\vec{P}_{\vec{R}}\cdot\dot{\vec{R}}+%
  \vec{p}_{\vec{r}}\cdot\dot{\vec{r}})-\frac{e^2}{r}.
\]

Solving equations \ref{momento del centro de masas} and
\ref{momento del movimiento interno} for the velocities we find
\begin{equation}\label{velocidad del centro de masas}
  \dot{\vec{R}}=\frac{\mu\left(1+\frac{2e^2}{Mc^2r}
  \right)\vec{P}_{\vec{R}}+\frac{K_Le^2}{c^2r}\vec{p}_{\vec{r}}}{\Delta},
\end{equation}
and
\begin{equation}\label{velocidad del movimiento interno}
  \dot{\vec{r}}=\frac{  M\left(1-\frac{2e^2}{Mc^2r}  \right)\vec{p}_{\vec{r}}+\frac{K_L e^2}{c^2
  r}\vec{P}_{\vec{R}}
}{\Delta},
 \end{equation}
where
\[
\Delta=
M\mu\left(1-\frac{4e^4}{M^2c^4r^2}\right)-\frac{K_L^2e^4}{c^4r^2}=
m_pm_e-\frac{e^4}{c^4r^2}.
\]
Now we are ready to write the Hamilton's Function
\begin{equation}\label{funcion de hamilton en el centro de masas}
  H=\frac{1}{2}\frac{
    \left(1+\frac{2e^2}{Mc^2r}\right)\frac{\vec{P}_{\vec{R}}^{\
    2}}{M}+\left(1-\frac{2e^2}{Mc^2r}\right)\frac{\vec{p}_{\vec{r}}^{\
    2}}{\mu}+\frac{2K_Le^2}{m_pm_ec^2r}\vec{P}_{\vec{R}}\cdot\vec{p}_{\vec{r}}
  }{1-\frac{e^4}{m_pm_ec^4r^2}}-\frac{e^2}{r}
\end{equation}

\end{document}